\documentclass{PoS}

\title{Bottomonium Spectroscopy at CLEO}

\ShortTitle{Bottomonium Spectroscopy at CLEO}

\author{\speaker{Kamal K. Seth}\\ 
        Northwestern University, Evanston, IL 60208, USA\\
        E-mail: \email{kseth@northwestern.edu}}

\abstract{Results of the latest CLEO contribution to bottomonium spectroscopy is presented, the confirmation of the $\eta_b(^1S_0)$ ground state of bottomonium in the radiative decay $\Upsilon(3S)\to\gamma\eta_b$. The bottomonium hyperfine splitting is determined to be $\Delta M_{hf}(1S)=68.5\pm6.6\pm2.0$~MeV and the branching fraction $\mathcal{B}(\Upsilon(3S)\to\gamma\eta_b)=(7.1\pm1.8\pm1.1)\times10^{-4}$.  These results are in good agreement with those reported by BaBar.}

\FullConference{12th International Conference on B-Physics at Hadron Machines - BEAUTY 2009\\
		 September 07 - 11 2009\\
		 Heidelberg, Germany}

\begin{document}

\section{Introduction}

The $\Upsilon(1^3S_1)$ state of bottomonium was discovered in 1972~\cite{ups-disc}.  However, its spin-singlet partner, $\eta_b(1^1S_0)$, the ground state of bottomonium, eluded all attempts for identification for 36 years.  These included unsuccessful attempts by CUSB and CLEO at Cornell, and ALEPH and DELPHI at CERN.  In particular, CLEO searched for $\eta_b$ in radiative decays of $\Upsilon(3S)$ and $\Upsilon(2S)$ and reported upper limits for the branching fractions $\mathcal{B}(\Upsilon(2S,3S)\to\gamma\eta_b)$~\cite{ups-cleo}.  In July 2008 the first successful observation of $\eta_b$ was reported by BaBar~\cite{ups-babar}, and in the present talk I am describing  the independent confirmation of this observation by CLEO~\cite{ups-cleo-new}.

To provide a perspective, the bottomonium spectrum is shown in Fig.~1.  We also note that CLEO III acquired data at $\Upsilon(1S,2S,3S)$ with luminosities of $\sim1.1$, 1.2 and 1.2 fb$^{-1}$, respectively, whereas the corresponding BaBar luminosities were 14.45~fb$^{-1}$ and 30.2~fb$^{-1}$ at $\Upsilon(2S)$ and $\Upsilon(3S)$, respectively.

The Babar observation of $\eta_b$ in the analysis of their data for 109 million $\Upsilon(3S)$ in the reaction $\Upsilon(3S)\to\gamma\eta_b(1S)$ is shown in Fig.~2.  BaBar's success in identifying $\eta_b$ owed not just to their large data set but also to achieving a large reduction in background by the using a cut on the thrust angle, the angle between the signal photon and the thrust vector of the rest of the event.  By making a cut at $|\cos\theta_T|\ge0.7$ they achieved a nearly factor three reduction in the continuum background at the cost of sacrificing $\sim30\%$ of the $\eta_b$ signal.

\begin{figure}[b]
\begin{center}
\includegraphics[width=4.in]{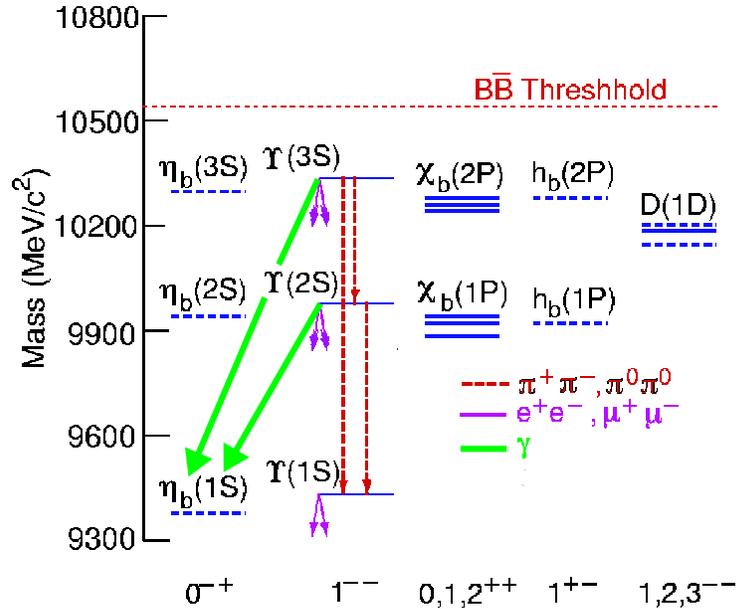}
\end{center}
\caption{Spectrum of the bound states of the ($b\bar{b}$) Upsilon family.}
\end{figure}

\begin{figure}
\begin{center}
\includegraphics[width=2.4in]{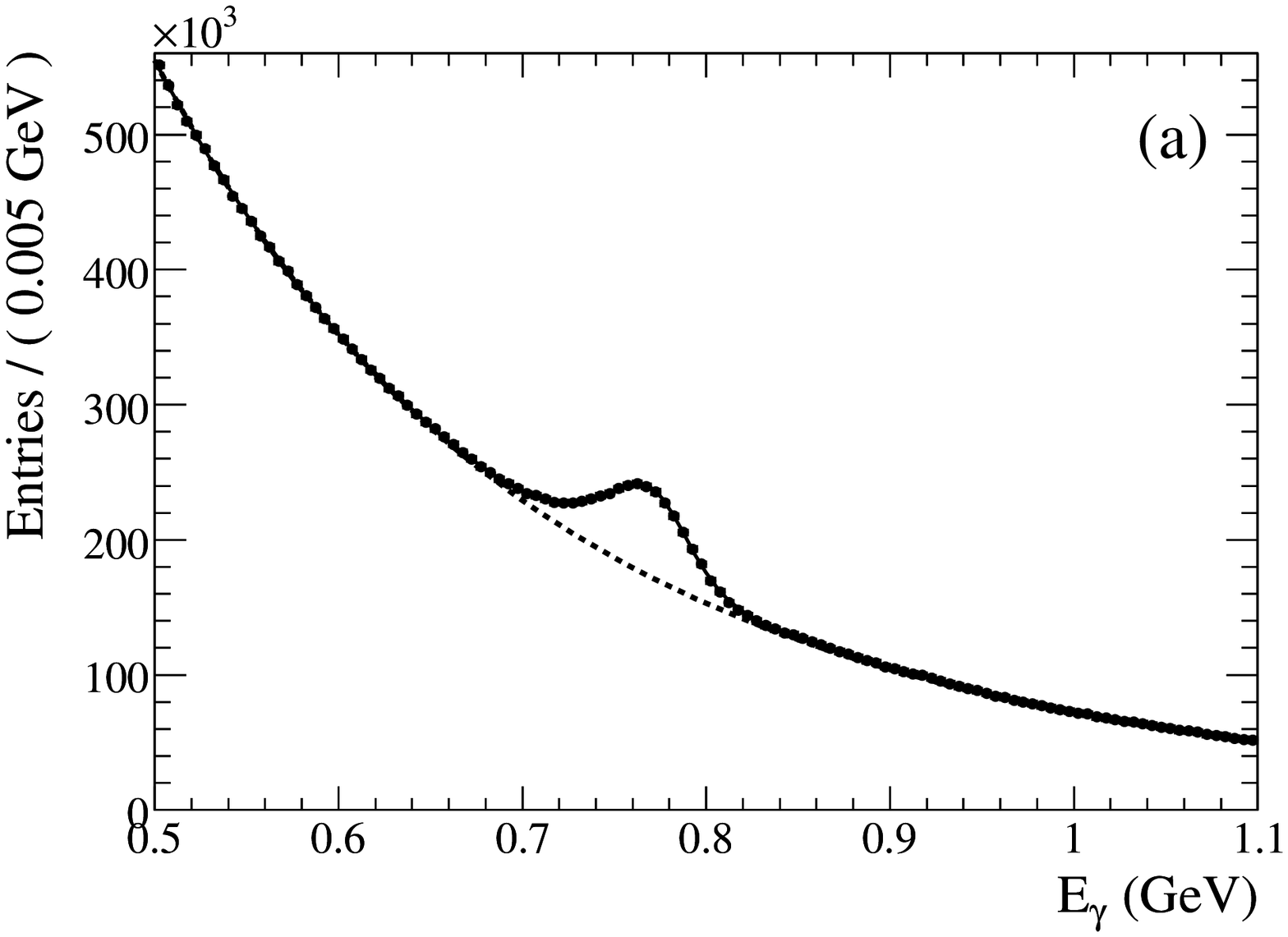}
\includegraphics[width=2.5in]{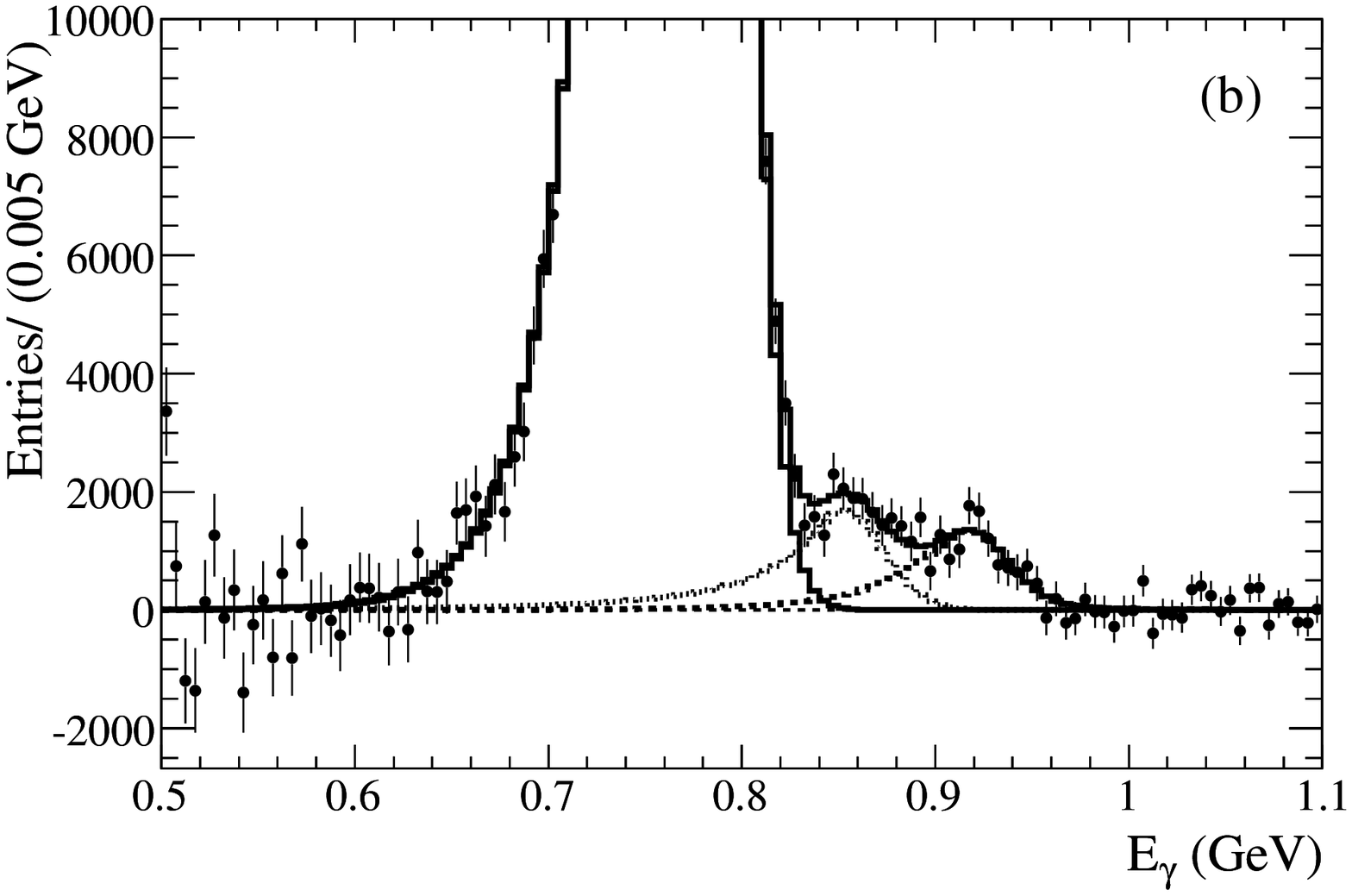}
\end{center}
\caption{BaBar results for the observation of $\eta_b$.  (Left) The gross features of the inclusive photon spctrum.  (Right) The background subtracted photon spectrum.  The peaks, from left to right, are from $\chi_{bJ}$, ISR, and $\eta_b$.}
\end{figure}

In order to succeed in identifying the $\eta_b$ signal with a factor 20 smaller data set (CLEO's 5.9 million $\Upsilon(3S)$ versus BaBar's 109 million $\Upsilon(3S)$) we had to make several improvements over BaBar's analysis procedure.

Fig.~3 provides the perspective for data analysis and the challenges involved in identifying $\eta_b$ in presence of the huge continuum backgrounds in the inclusive photon spectra for $\Upsilon(3S)$ and $\Upsilon(2S)$ radiative decays. In $\Upsilon(3S)$ data, the only visible peak is due to the unresolved transitions $\Upsilon(3S)\to\gamma\chi_{bJ},~\chi_{bJ}\to\gamma\Upsilon(1S)$ ($J=0,1,2$) at $E_\gamma\sim750$~MeV.  On the high energy tail of the $\chi_{bJ}$ peak lie the much weaker (factors$>20$) transitions, the ISR transition $\Upsilon(3S)\to\gamma_{ISR}\Upsilon(1S)$ at $\sim860$~MeV and $\Upsilon(3S)\to\gamma\eta_b(1S)$ at $\sim920$~MeV.  In order to identify these very weak transitions three things are necessary: it is necessary to minimize the effects of the background and study its parameterization carefully, it is necessary to have an accurate parameterization of the shapes of the photon peaks whose tails overlap, and it is necessary to preserve the full statistics of the data by not rejecting any part of it.

\begin{figure}
\begin{center}
\includegraphics[width=2.6in]{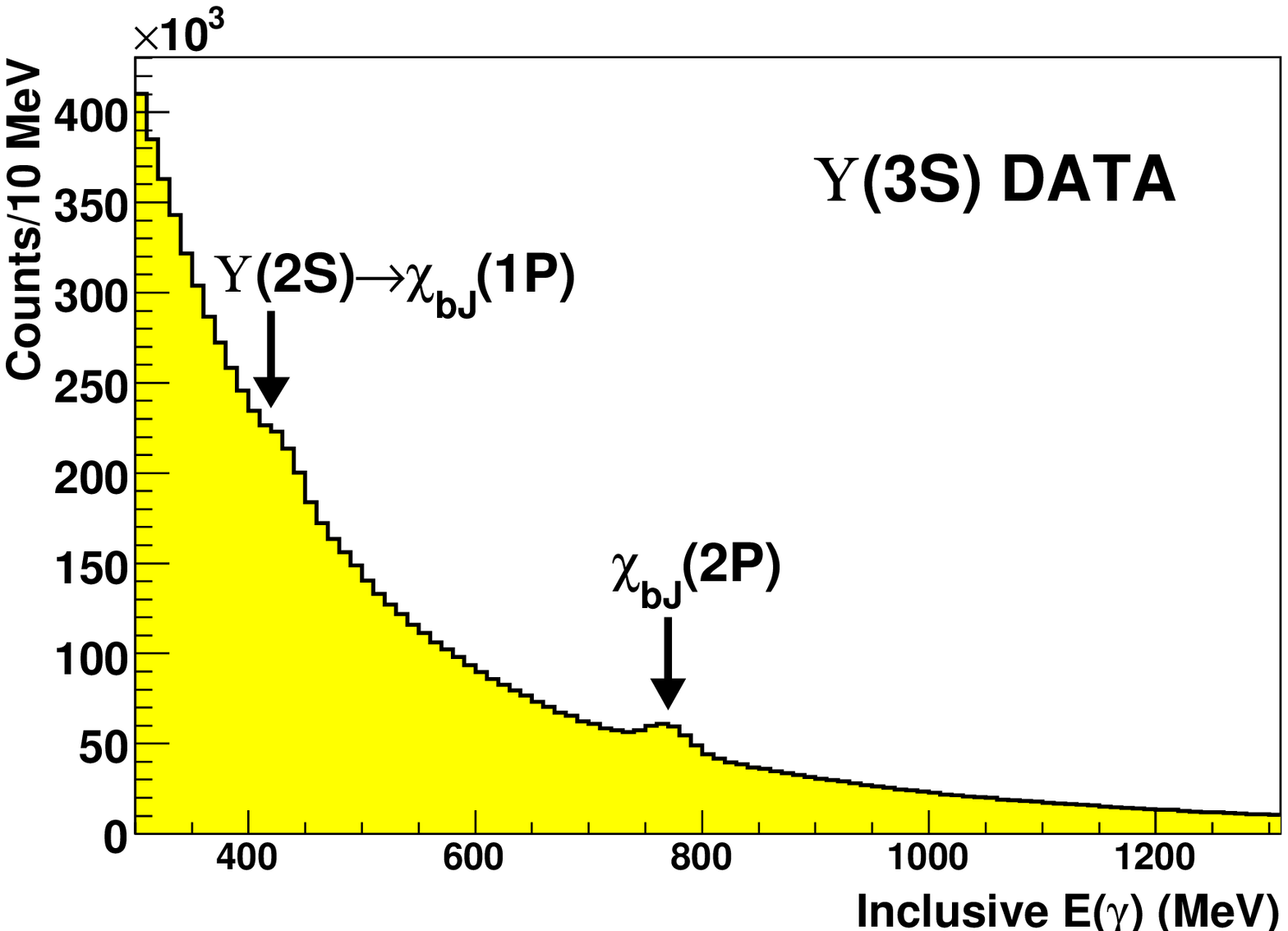}
\includegraphics[width=2.6in]{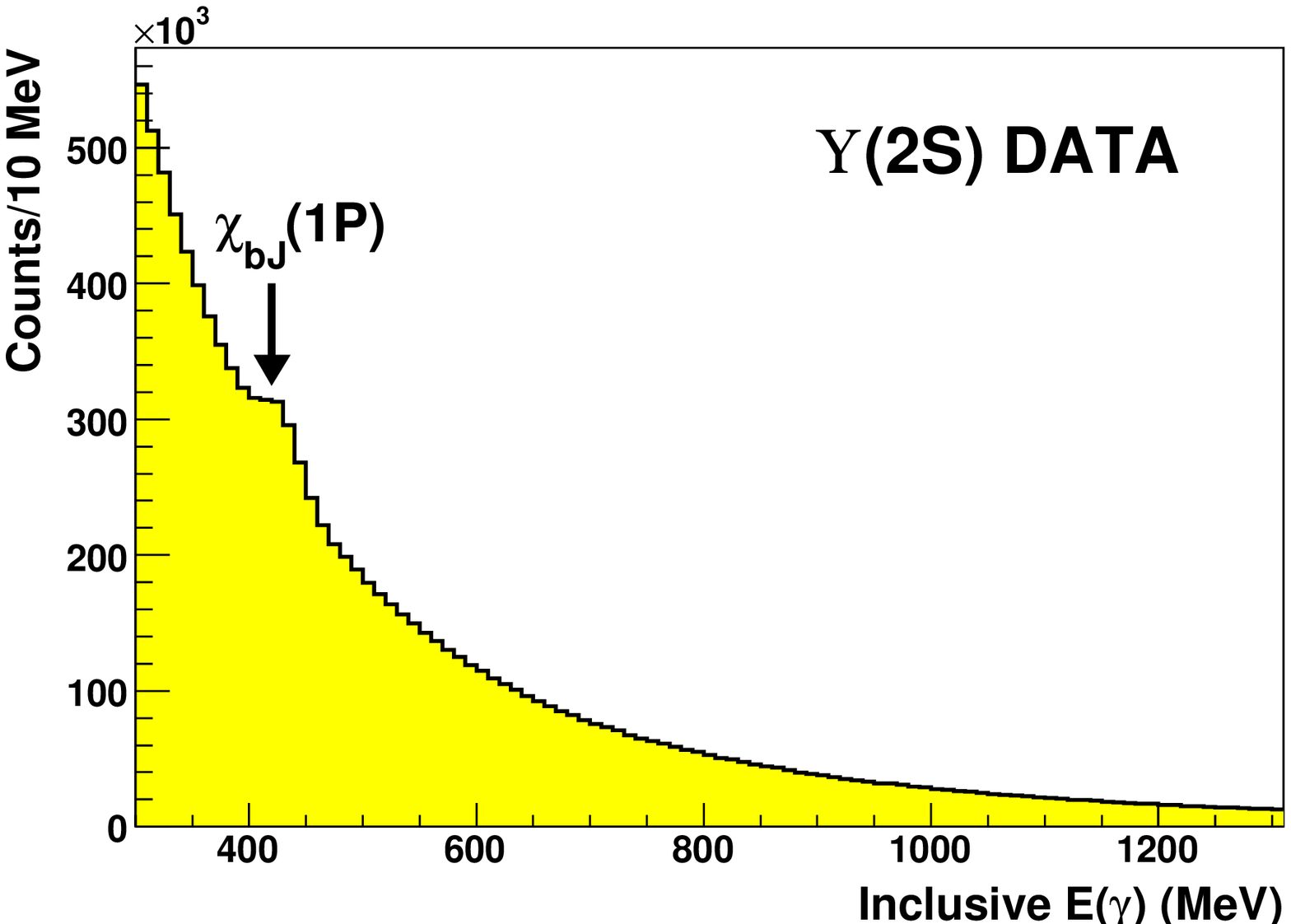}
\end{center}
\caption{CLEO spectra illustrating the gross features of the inclusive photon spectra for (left) $\Upsilon(3S)$ decay, and (right) $\Upsilon(2S)$.}
\end{figure}

\textbf{Photon Line Shapes:} As is well known, photon lines in an electromagnetic calorimeter acquire low energy tails which are usually parameterized in terms of the Crystal Ball parameters; $\sigma$, the Gaussian width, $\alpha$, the matching point of the tail, and $n$, the rate of fall of the tail.  An accurate determination of the tail parameters can only be done from background-free photon lines.  We do so in two independent ways.  In one method we use the observed shapes of the background-free photons of a given energy from radiative Bhabhas, and in the other we use the shape of the background-free peaks of the exclusive decays $\chi_{b1}(2P,1P)\to\gamma\Upsilon(1S),~\Upsilon(1S)\to l^+l^-$.  The two methods give consistent values of the parameters which are fixed in the subsequent analyses.

\begin{figure}
\begin{center}

\includegraphics[width=2.9in]{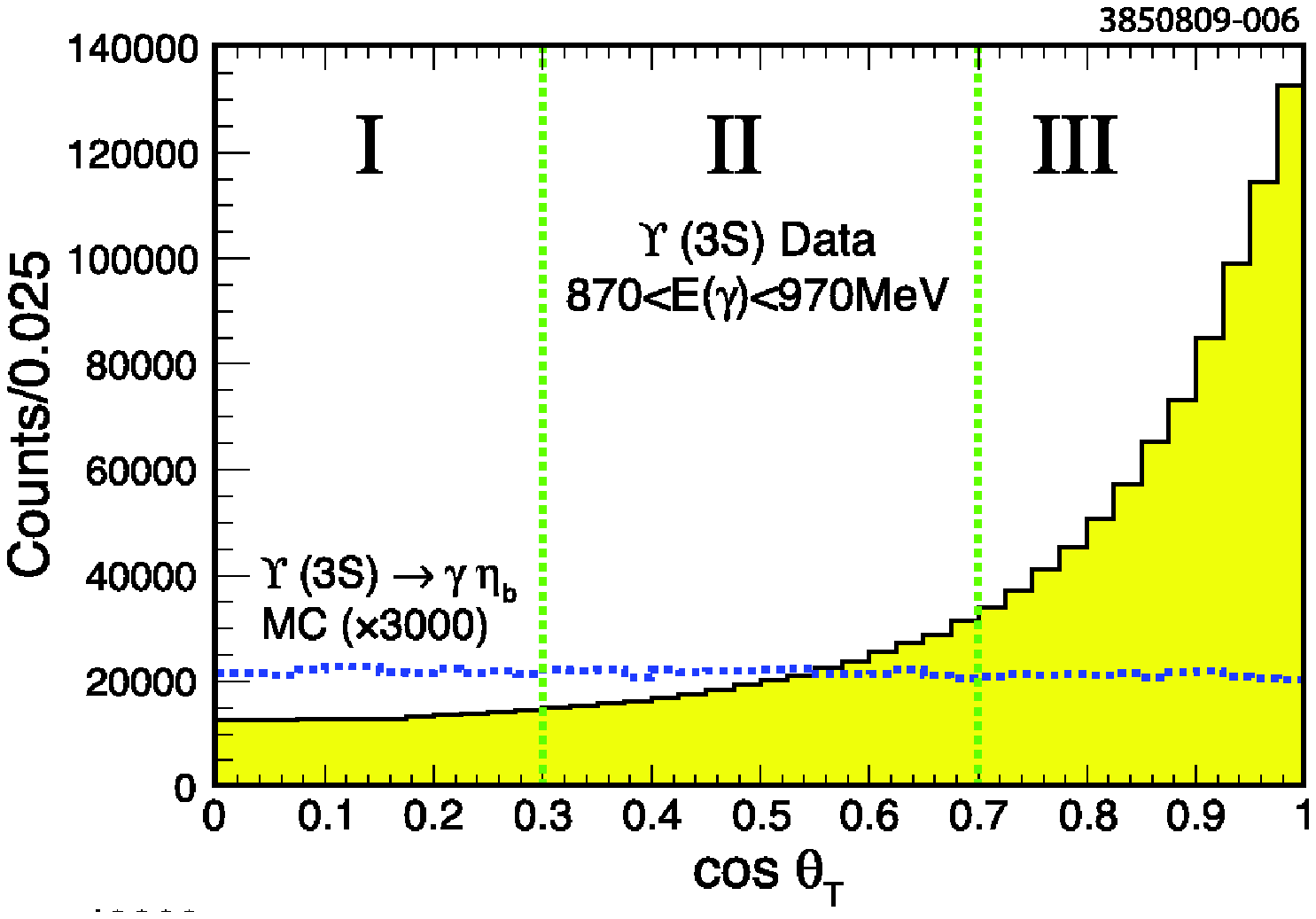}
\includegraphics[width=2.9in]{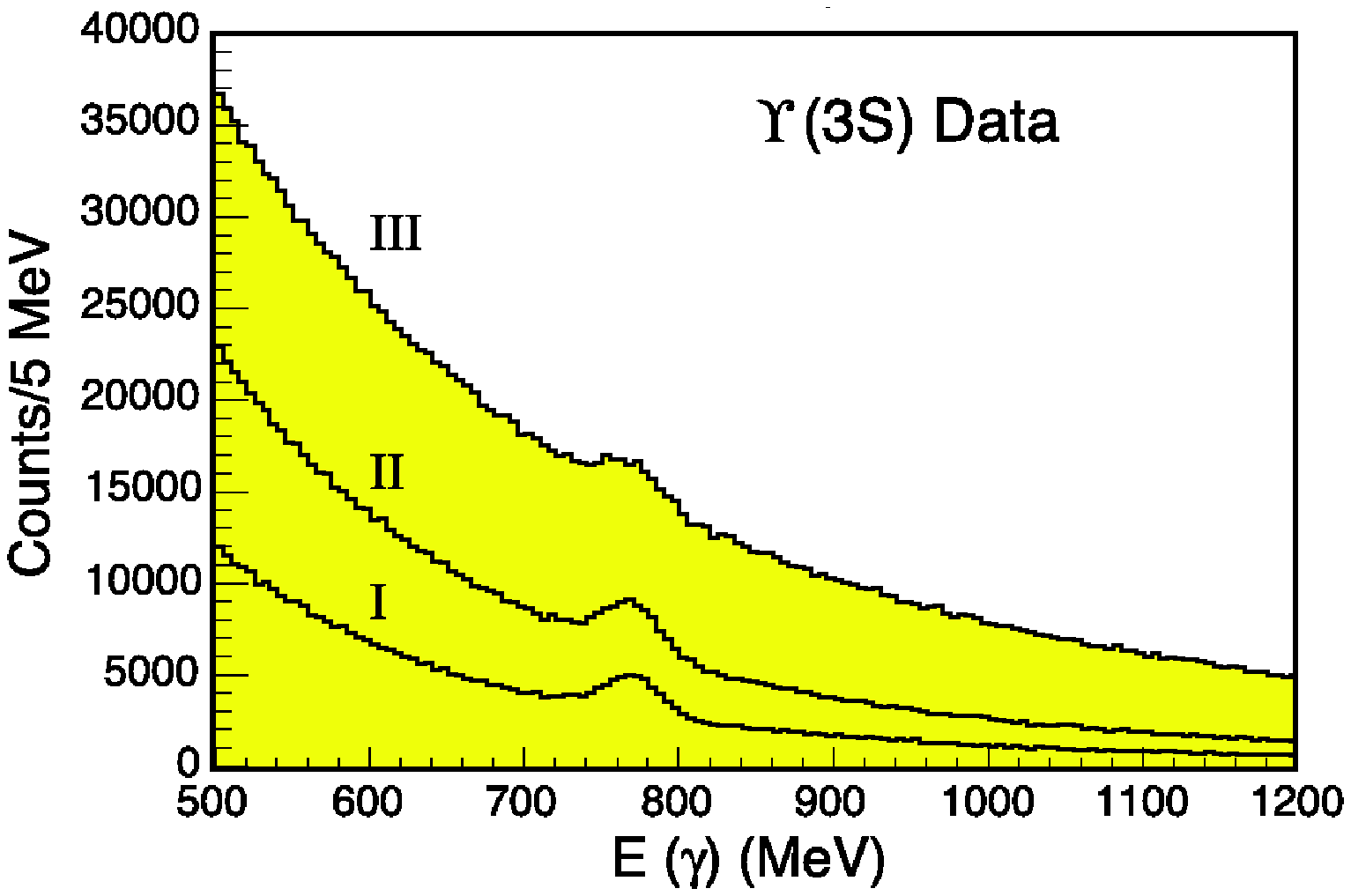}
\end{center}
\caption{(Left) Thrust angle $|\cos\theta_T|$ distributions for $\Upsilon(3S)$ data and the expected MC distribution for the $\eta_b$ signal.  (Right) Illustrating the different signal/background ratios in the three regions of $|\cos\theta_T|$.}
\end{figure}

\textbf{Background Parameterization:} We find that the fits to the continuum background are the crucial determinant in the results for the weak $\eta_b$ peak.  We also find that equally good fits to the background  can be obtained with different parameterizations, range of fits, and methods of binning the data. We have made a large number of background fits (several hundred) to the data in each of three bins of $|\cos\theta_T|$, I:~$|\cos\theta_T|=0-0.3$,  II:~$|\cos\theta_T|=0.3-0.7$,  III:~$|\cos\theta_T|=0.7-1.0$, 
using exponential polynomials of various orders (2,3,4), in various energy regions (500--1340~MeV), and with linear and log binning of the data. 

The average results for $E_\gamma(\eta_b)$, $\mathcal{B}(\Upsilon(nS)\to\gamma\eta_b)$, and significance for all the good fits (CL$>10\%$) were considered as our final results, and their r.m.s. variations were taken as measures of the systematic uncertainties in the results, $\pm1$~MeV in $E_\gamma$, $\pm10\%$ in $\mathcal{B}(\eta_b)$, and $\pm0.4\sigma$ in significance.

\textbf{ISR Peak:} The energy of the ISR photon peak in $\Upsilon(nS)\to\gamma_\mathrm{ISR}\Upsilon(1S)$ is accurately known, and was fixed.  The yield of the ISR peak was estimated by extrapolating the observed yield in CLEO data taken at $\Upsilon(4S)$.  It was then fixed to this value.

\textbf{Method of Joint Analysis of Data in Three $|\cos\theta_T|$ Bins:} As illustrated in the top panel of Fig.~4, the $|\cos\theta_T|$ distribution for the background--dominated data is peaked in the forward direction, $|\cos\theta_T|\approx1$, whereas for the $\eta_b$ it is expected to be uniform. As a result, the data in the three different regions of $|\cos\theta_T|$ have different ratios of signal/background as shown in the bottom plot.

Unlike BaBar, we do not cut off the $|\cos\theta_T|>0.7$ region.  Instead, we let each region contribute to the total result weighted by its individual signal-to-background.  Since none of the data are rejected, we preserve full statistics, and are free of uncertainties in alternate implementations of the thrust cuts.  We have analyzed our data by the joint fit method, and for comparison purposes also with $|\cos\theta_T|<0.7$.  We find that the joint fit method enhances the significance of the $\eta_b$ identification by $\sim1\sigma$.

\subsection{Results for $\Upsilon(3S)\to\gamma\eta_b(1S)$}

A representative fit from the joint fit analysis of the three regions of the thrust angle is shown in Fig.~5.  As is shown there, the ISR and $\eta_b$ peaks are clearly visible in the region~I ($|\cos\theta_T|=0-0.3$), less so in region~II  ($|\cos\theta_T|=0.3-0.7$), and difficult to discern in region~III ($|\cos\theta_T|=0.7-1.0$).  Yet, all three regions contribute to the identification of $\eta_b$.  The fit has $N(\eta_b)=2311\pm546$~counts, $E_\gamma(\eta_b)=918.6\pm6.0$(stat)~MeV, which corresponds to the hyperfine splitting of $68.5\pm6.6$~MeV, and~$\mathcal{B}(\Upsilon(3S)\to\gamma\eta_b)=(7.1\pm1.8(\mathrm{stat}))\times10^{-4}$.

We have made a very conservative evaluation of systematic uncertainties in our data.  These are listed in Table~1.  We note that in our analysis we have assumed $\Gamma(\eta_b)=10$~MeV.  However, we find that the branching fraction depends linearly on the assumed width, $\Gamma(\eta_b)$, as~$\mathcal{B}(\Upsilon(3S)\to\gamma\eta_b)\times10^4=5.8+0.13(\Gamma(\eta_b)\mathrm{~in~MeV})$.  Our final results are presented in Table~II.

\subsection{Results for $\Upsilon(2S)\to\gamma\eta_b(1S)$}

We have analyzed our data for $\Upsilon(2S)$ radiative decay in exactly the same manner as for $\Upsilon(3S)$.  However, because the continuum background in the vicinity of $\Upsilon(2S)\to\gamma\eta_b(1S)$ transition, expected at $E_\gamma\approx611$~MeV, is nearly six times larger than the corresponding background in $\Upsilon(3S)$ decay (see Fig.~3), the $\eta_b$ signal is not observed in either of the three $|\cos\theta_T|$ bins.  This is illustrated in Fig.~6.  As a result it was only possible to establish an upper limit for the branching fraction for the transition, $\mathcal{B}(\Upsilon(2S)\to\gamma\eta_b)<8.4\times10^{-4}$ at the 90\% confidence level.

In Table II, we summarize our final results.  For comparison the corresponding results of BaBar are also listed.  The two are in agreement.  In Table~II we also list the theoretical predictions for hyperfine splitting and branching fractions.  The pQCD-based predictions for both vary between wide limits.  Recently, predictions for $\Delta M_{hf}(1S)_{b\bar{b}}$ have also become available from Lattice calculations.  The results of three of them \cite{lqcd} are also listed in Table~II.  The predictions are in general agreement with the experimental results.

\begin{figure}
\begin{center}
\includegraphics[width=3.5in]{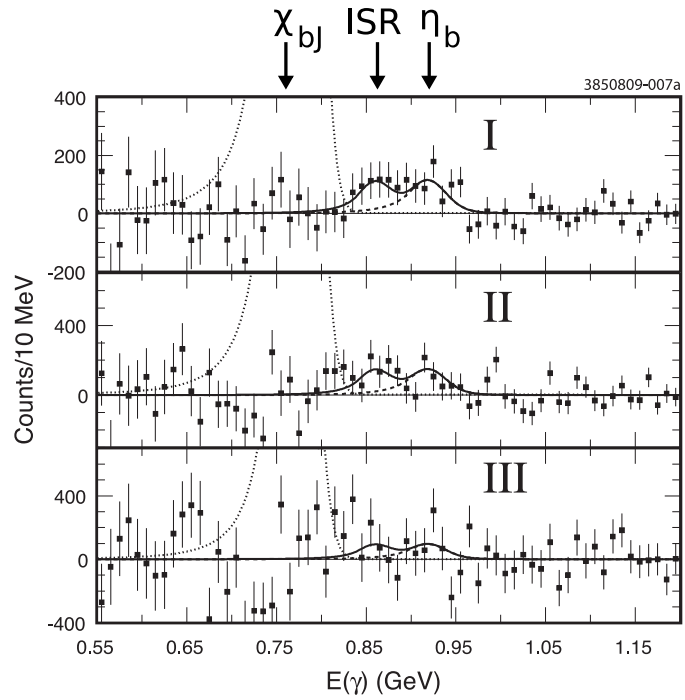}
\end{center}
\caption{Illustrating joint fit results for the background subtracted spectra for $\Upsilon(3S)\to\gamma\eta_b(1S)$.}
\end{figure}

\begin{figure}
\begin{center}
\includegraphics[width=3.5in]{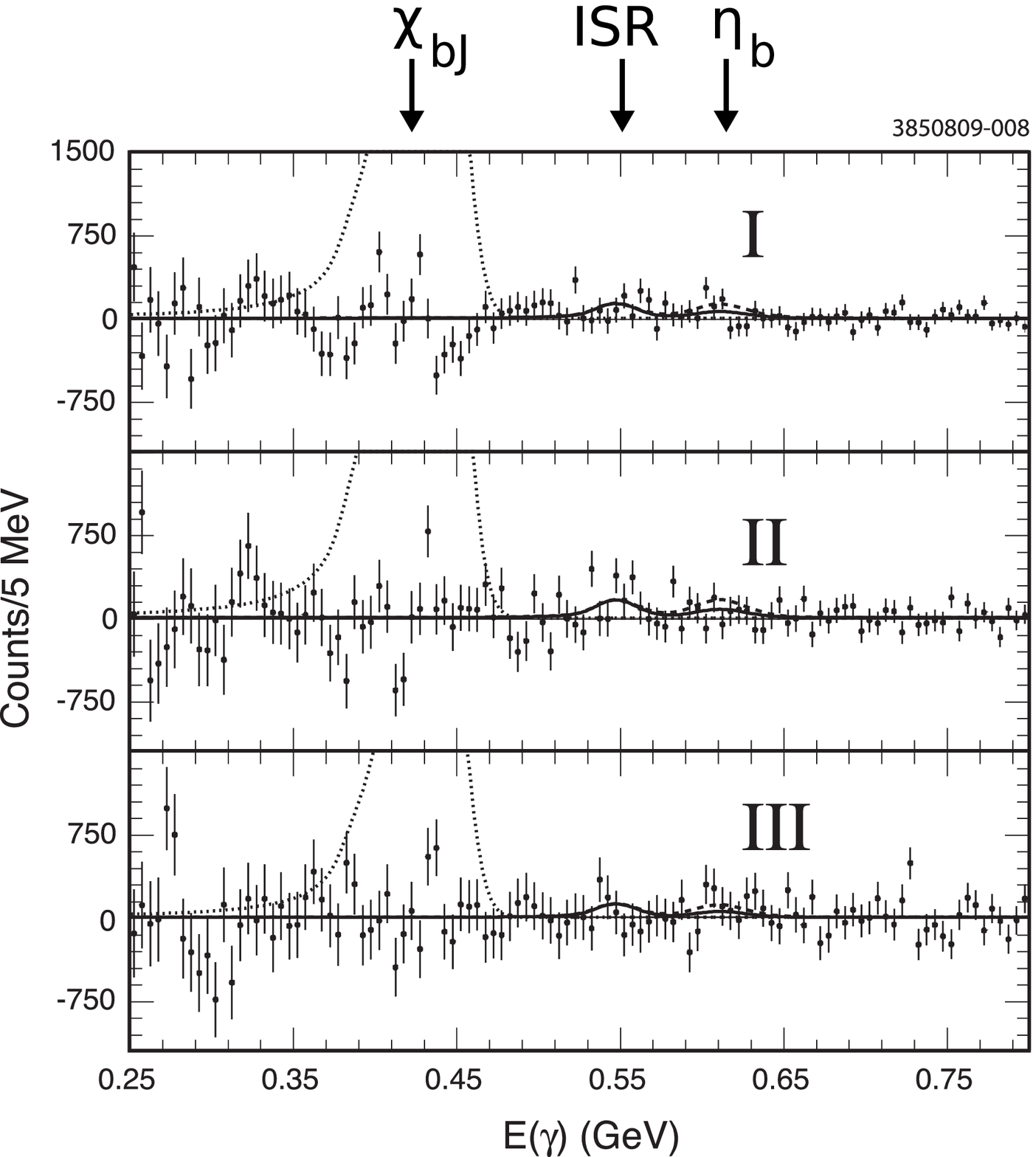}
\end{center}
\caption{Illustrating joint fit results for the background subtracted spectra for $\Upsilon(2S)\to\gamma\eta_b(1S)$.}
\end{figure}

\begin{table}
\begin{center}
\begin{tabular}{lcc}
\hline \hline
 & \multicolumn{2}{c}{\textbf{Uncertainty in}} \\
\textbf{Source} & $E_\gamma$ \textbf{(MeV)} & $\mathcal{B}(\Upsilon\to\gamma\eta_b)$ \\
\hline
Background  (fn, range, binning)   & $\pm1.0$ & $\pm10\%^\dagger$ \\
Photon Energy Calibration & $\pm1.2^*$ & --- \\
Photon Energy Resolution  & $\pm0.3$ & $\pm2\%$ \\
CB and $\chi_{bJ}(2P)$ Parameters   &    $\pm0.7$ & $\pm8\%^\dagger$ \\
ISR Yield & $\pm0.4$ & $\pm3\%$ \\
Photon Reconstruction & --- & $\pm2\%$ \\
$N(\Upsilon(3S))$ & --- & $\pm2\%$ \\
MC Efficiency  & --- & $\pm7\%$ \\
\hline
\textbf{Total} & $\pm1.8$ & $\pm15$\% \\
\hline \hline
\end{tabular}
\end{center}
\begin{small}
\begin{itemize}
\item[$^*$] Our ISR photon energy agrees with the expected energy within 0.3~MeV.\\
Our $\chi_{bJ}(2P)$ centroid energy agrees with the expected energy within 0.3~MeV.\vspace*{-5pt}
\item[$^\dagger$] Despite very detailed studies of background and peak shape parameters, we assign these large uncertainties to be very conservative.
\end{itemize}
\end{small}
\caption{Systematic error contributions for $\Upsilon(3S)\to\gamma\eta_b$ photon energy and branching fraction.}
\end{table}

\begin{table}
\begin{center}
\begin{tabular}{llccc}
\hline
& & $\Delta M_{hf}(1S)_{b\bar{b}}$, (MeV) & $\mathcal{B}(\Upsilon(nS)\to\gamma\eta_b)\times10^4$ & significance \\
\hline
$\Upsilon(3S)\to\gamma\eta_b$ & (CLEO) & $68.5\pm6.6\pm2.0$ & $7.1\pm1.8\pm1.1$ & $4\sigma$ \\
                              & (BaBar)& $71.4^{+3.1}_{-2.3}\pm2.7$ & $4.8\pm0.5\pm0.6$ & $\ge10\sigma$\\
$\Upsilon(2S)\to\gamma\eta_b$ & (CLEO) & --- & $<8.4$ (90\% CL) & --- \\
                              & (BaBar)& $67.4^{+4.8}_{-4.6}\pm2.0$ & $4.2^{+1.1}_{-1.0}\pm0.9$ & $3.5\sigma$ \\
\hline
\multicolumn{2}{l}{Lattice  (UKQCD+HPQCD)}  & $61\pm14$ \\
        & (TWQCD) & $70\pm5$ \\
        & (Ehmann) & $37\pm8$ \\
\multicolumn{2}{l}{pQCD (various)}    & $35-100$ & $0.05-25$ ($\Upsilon(3S)$) \\
 & & & $0.05-15$ ($\Upsilon(2S)$) \\
\hline
\end{tabular}
\end{center}
\caption{Summary of $\eta_b$ results and theoretical predictions.}
\end{table}

\section{Other $\Upsilon(nS)$ Results from CLEO}

Because of time constraints I have confined myself to the latest CLEO result, the confirmation of the $\eta_b$ discovery.  Let me, however, mention six recently published CLEO papers on physics from the $\Upsilon(nS)$ data.
\begin{enumerate}
\item ``Observation of $\Upsilon(2S)\to\eta\Upsilon(1S)$ and Search for Related Transitions'', \emph{Phys. Rev. Lett.} \textbf{101}, 192001 (2008).

\item ``Improved Measurement of Branching Fractions for $\pi\pi$ Transitions among $\Upsilon(nS)$ States'', \emph{Phys. Rev.} \textbf{D 79}, 011103(R) (2009).

\item ''Inclusive $\chi_{bJ}(nP)$ Decays to $D^0X$'', \emph{Phys. Rev.} \textbf{D 78}, 092007 (2008).

\item ''Observation of $\chi_{b}(1P_J,2P_J)$ Decays to Light Hadrons'', \emph{Phys. Rev.} \textbf{D 78}, 091103(R) (2008).

\item  ``Search for Very Light CP--odd Higgs in Radiative Decays of $\Upsilon(1S)$'', \emph{Phys. Rev. Lett.} \textbf{101}, 151802 (2008).

\item  ``Search for Lepton Flavor Violation in Upsilon Decays'',  \emph{Phys. Rev. Lett.} \textbf{101}, 201601 (2008).

\end{enumerate}

\end{document}